\documentclass[a4paper,12pt]{article}
\usepackage[]{}

\date{}

\def\build#1_#2^#3{\mathrel{\mathop{\kern 0pt#1}\limits_{#2}^{#3}}}
\begin{document}

\title{A POSSIBLE TWO-COMPONENT STRUCTURE OF THE NON-PERTURBATIVE 
POMERON}

\vfil

\author{ Pierre GAURON and Basarab NICOLESCU \\ 
\\
	LPNHE\thanks{Unit\'e de Recherche des Universit\'es Paris 6 et 
		Paris 7, Associ\'ee au CNRS} - LPTPE, Universit\'e Pierre et Marie 
		Curie,\\ 4, Place Jussieu, 75252 Paris Cedex 05, France\\
                 gauron@in2p3.fr, nicolesc@lpnp66.in2p3.fr}

\vfil

\maketitle
\noindent {\bf Abstract}

We propose a QCD-inspired two-component Pomeron form which gives an 
excellent description of the $pp,\ \pi p,\ Kp,\ \gamma p$ and $\gamma\gamma$
total cross-sections. Our fit has a better $\chi^{2}/dof$ for a smaller   
number of parameters as compared with the PDG fit. Our 2-Pomeron form is
fully compatible with weak Regge exchange-degeneracy, universality, 
Regge factorization and the generalized vector dominance model.\\

\vspace{2.5in}

LPNHE 00-02\hspace{3.0in}April 2000

\newpage

40 years after its introduction \cite{1} and in spite of very 
important advances in QCD, the Pomeron remains an open problem. 
In particular, the non-perturbative structure of the Pomeron is 
still controversial.

The most popular model of the non-perturbative Pomeron is, of course, 
the Donnachie-Landshoff (DL) model \cite{2}. The total cross-sections 
for $pp$ and $\bar pp$ scattering are parametrized in terms of five
parameters~:
\begin{equation}
	\sigma_{pp}=X_{pp}s^\varepsilon+Y_{pp}s^{-\eta}
	\label{1}
\end{equation}
\begin{equation}
	\sigma_{\bar pp}=X_{pp}s^\varepsilon+Y_{\bar pp}s^{-\eta}
	\label{2}
\end{equation}
where
\begin{equation}
	\varepsilon=\alpha_{P}(0)-1
	\label{3}
\end{equation}
and
\begin{equation}
	\eta=1-\alpha_{R}(0)\ ;
	\label{4}
\end{equation}
$\alpha_{P}(0)$ is the Pomeron-intercept, $\alpha_{R}(0)$ is the 
effective non-leading exchange-degenerate Regge intercept and $X,\ Y$ 
the corresponding Regge residues. An overall scale factor $s_{0}=1$ 
GeV$^{2}$ is implicitely present in eqs. (\ref{1})-(\ref{2}). The 
key-parameters $\varepsilon$ and $\alpha_{R}(0)$ have the following 
values~:
\begin{equation}
	\varepsilon=0.0808
	\label{5}
\end{equation}
and
\begin{equation}
	\alpha_{R}(0)=0.5475.
	\label{6}
\end{equation}
The $pp$ data are well reproduced. It was therefore tempting to use 
the DL form to the simultaneous study of all existing total 
cross-sections. It is precisely what was done by PDG in the last 
edition of the "Review of Particle Physics" \cite{3}, \cite{4}.

The total cross-sections $\sigma$ are parametrized in refs. 3 and 4  
in the variant of a non-exchange-degenerate DL form~:
\begin{equation}
	\sigma_{AB}=X_{AB}s^\varepsilon +Y_{1AB}s^{-\eta_{1}}
	-Y_{2AB}s^{-\eta_{2}},
	\label{7}
\end{equation}
\begin{equation}
	\sigma_{\bar AB}=X_{AB}s^\varepsilon +Y_{1AB}s^{-\eta_{1}}
	+Y_{2AB}s^{-\eta_{2}},
	\label{8}
\end{equation}
where
\begin{equation}
	\eta_{1}=1-\alpha_{R_{+}}(0),\quad \eta_{2}=1-\alpha_{R_{-}}(0),
	\label{9}
\end{equation}
$\alpha_{R_{+}}(0)$ and $\alpha_{R_{-}}(0)$ being the Regge intercepts
of the non-leading Regge trajectory $R_{+}$ in the even-under-crossing
amplitude and $R_{-}$ in the
odd-under-crossing amplitude respectively. $X,\ Y_{1},\ Y_{2}$ are the
corresponding Regge residues. There are 16 parameters for fitting 271
experimental points involving 8 reactions~: $\bar pp,\ pp,\ \pi^\pm p,\
K^\pm p,\ \gamma p$ and $\gamma\gamma$. The overall $\chi^{2}$ is
excellent~: $\chi^{2}/dof=0.93$\footnote{A bigger value
$\chi^{2}/dof=1.02$, corresponding to 383 experimental points and
$\varepsilon =0.0933$, is quoted in table 1 of ref. 4  
because real parts are also included in the 
respective fits (see text for a discussion of this option).}.
The key-parameter $\varepsilon$ has now the value 0.0900.

The problem with the form of refs. 3 and 4 is the bad violation of the weak 
exchange-degeneracy (i.e. $\alpha_{R_{+}}(0)=\alpha_{R_{-}}(0))$, namely
$$
\alpha_{R_{+}}(0)-\alpha_{R_{-}}(0)\simeq 0.2.
$$
However, the \textsl{masses of the resonances}, as published in the
"Review of Particle Physics" \cite{5}, clearly indicate that the weak 
exchange-degeneracy is respected. As seen from fig. 1a) the 10 
resonances belonging to the 4 different $I^{G}(J^{PC})$ families
$\rho-\omega-f_{2}-a_{2}$ are compatible with a \textsl{unique} linear 
exchange-degenerate Regge trajectory
\begin{equation}
	\alpha(t)=\alpha(0)+\alpha't
	\label{10}
\end{equation}
with
\begin{equation}
	\alpha(0)=0.48
	\label{11}
\end{equation}
and
\begin{equation}
	\alpha'=0.88\ (GeV/c)^{-2}.
	\label{12}
\end{equation}
The numerical values (\ref{11})-(\ref{12}) are extracted just by 
plugging in (\ref{10}) the masses and the spins of $\rho_{1}(770)$ and 
$\rho_{3}(1690)$ resonances.

Remarkably enough, the \textsl{same} $\alpha(0)$ value (\ref{11}) is 
compatible with the $\Delta\sigma$ data for the total cross-section 
differences
\begin{equation}
	\Delta\sigma_{AB}\equiv \sigma_{\bar AB}-\sigma_{AB}=2Y_{2AB}
	s^{\alpha_{R_{-}}(0)-1}
	\label{13}
\end{equation}
or
\begin{equation}
	\mbox{ln}\ [s\Delta\sigma_{AB}]=\mbox{ln}\ (2Y_{2AB})+
	\alpha_{R_{-}}(0)\mbox{ln}\ s.
	\label{14}
\end{equation}
The $\Delta\sigma$ data for $pp,\ Kp$ and $\pi p$ and $\sqrt{s}
\build >_{\sim}^{} 6$ GeV \cite{6} shown in the log-log plot of fig. 1b) 
are all compatible with the straight lines of eq. \ref{14}~, the slopes 
of which are 
precisely given by the $\alpha_{R_{-}}(0)$ value of eq. (\ref{11}).  

These  indications in favour of the weak exchange-degeneracy, coming 
both from the resonance and scattering region, is too striking to be a 
mere coincidence. One can therefore wonder if something is inadequate 
in the parametrization (7-8).

A first problem can come from the fact that the ratio $\rho(s,t=0)=\mbox{Re}
F(s,t=0)/\mbox{Im}F(s,t=0)$  has been included into the PDG fit together with
the total cross-sections. As it is known, the determination of this $\rho$ 
parameter is semi-theoretical~: its value is obtained through an extrapolation
of the elastic amplitude to t=0 using a theoretical model. The result is very
sensitive to these theoretical assumptions (see, for example, \cite{7}). If we 
try to redo the minimization using the total cross-sections only, the
violation of the weak Regge exchange-degeneracy persists, as already noted in
\cite{8}.
In the following we will minimize with different analytic forms
but using the total cross-sections only. 

An important problem may come from the form of the Pomeron. The
non-perturbative Pomeron is certainly much more complex than a simple pole,
which violates the unitarity. It surely includes cuts associated 
with multiexchanges which restore unitarity.
We do not know the exact form of these complicate
singularities. But we can try to mimic them by a 2-component Pomeron, 
 a Pomeron built from two Regge singularities.

The perturbative Pomeron has also a complex form. The BFKL Pomeron is
not a simple pole but rather a complicate cut or an accumulation of poles
close to $J=1$. Also, very recently, detailed calculations in the perturbative 
QCD indicate, in fact, the existence of a 2-component Pomeron. Namely, in 
LLA, beside the BFKL Pomeron associated with 2-gluon exchange and 
corresponding to an intercept $\alpha_{p}^{2g}(0)>1$, one finds a new Pomeron 
associated with the 3-gluon exchange with $C=+1$ and corresponding to 
an intercept $\alpha_{p}^{3g}(0)=1$~; the 3-gluon Pomeron is 
exchange-degenerate with the 3-gluon $C=-1$ Odderon \cite{9}.

Inspired by these considerations, we 
explore in this paper the possibility of a 2-component Pomeron in the 
non-perturbative sector, namely a Pomeron built from two poles.
We propose the new analytic forms for 
the total cross-sections~:

\begin{equation}
	\begin{array}{rcl}
		\sigma_{pp}  & = & Z_{pp}+Xs^{\varepsilon}+(Y_{1}^{pp}-Y_{2}^{pp})
		s^{\alpha(0)-1}  \\
		\sigma_{\bar pp}  & = & Z_{pp}+Xs^{\varepsilon}+(Y_{1}^{pp}+Y_{2}^{pp})
		s^{\alpha(0)-1} \\
		\sigma_{\pi^{+}p} & = & Z_{\pi p}+Xs^{\varepsilon}+(Y_{1}^{\pi p}
		-Y_{2}^{\pi p})s^{\alpha(0)-1}   \\
		\sigma_{\pi^{-}p} & = & Z_{\pi p}+Xs^{\varepsilon}+(Y_{1}^{\pi p}
		+Y_{2}^{\pi p})s^{\alpha(0)-1}   \\
		\sigma_{K^{+}p} & = & Z_{Kp}+Xs^{\varepsilon}+(Y_{1}^{Kp}
		-Y_{2}^{Kp})s^{\alpha(0)-1}   \\
		\sigma_{K^{-}p} & = & Z_{Kp}+Xs^{\varepsilon}+(Y_{1}^{Kp}
		+Y_{2}^{Kp})s^{\alpha(0)-1}   \\
		\sigma_{\gamma p}  & = & \delta Z_{pp}+\delta Xs^{\varepsilon}+
		Y_{1}^{\gamma p}s^{\alpha(0)-1}   \\
 	    \sigma_{\gamma \gamma}  & = & \delta^{2} Z_{pp}+\delta^{2} 
	    Xs^{\varepsilon}+Y_{1}^{\gamma \gamma}s^{\alpha(0)-1}
	\end{array}
	\label{15}
\end{equation}
where $\alpha(0)$ is \textsl{fixed} to the value $\alpha(0)=0.48$ as 
given by resonance masses and a scale factor $s_{0}=1$ GeV$^{2}$ is 
implicitely supposed.

The Pomeron in eqs. (\ref{15}) has 2 components~: the $X$-component
corresponds to a Regge intercept bigger than 1 ($\varepsilon>0$) and
the $Z$-component corresponds to an intercept exactly localised at 1.
We suppose that the $X$-component is fully universal (its coupling is the same
in all hadron-hadron reactions, as well as the energy behaviour
$s^{\varepsilon}$), while the $Z$-component is not fully 
universal. It is tempting to interpret the $X$-component as the 
gluonic component of the non-perturbative Pomeron and the  $Z$-component
as its flavour-dependent non-perturbative component. It is interesting 
to note that the possibility of a fully universal Pomeron was already 
considered in literature \cite{10}.

Of course, there is no double counting. In the framework of the 
$S$-matrix Theory \cite{11}, there are 2 solutions of the Reggeon 
calculus~: a critical Pomeron with intercept equal to 1, leading 
asymptotically to a $(\mbox{ln}s)^{\eta} (\eta < 2)$ behaviour of the total
cross-sections, and a supercritical Pomeron with intercept higher than 
1, connected, at asymptotic energies, to the Froissart $\mbox{ln}^{2}s$ 
behaviour of the total cross-sections. In other words, the $X$ and $Z$ 
components correspond to 2 different Regge singularities.

Both components are supposed to obey the Regge factorization property.
This is realized via the $\delta$-parameter in eqs. (\ref{15}) for the  
$pp,\ \gamma p$ and$\gamma\gamma$ processes.

Finally, the secondary Regge pole of intercept $\alpha(0)$ corresponds 
to an exchange-degenerate trajectory, in agreement with our previous 
discussion.

The forms (\ref{15}) involve n=14 parameters (to 
be compared with the PDG value n=16).
The values of the parameters in eq. (\ref{15}) are given in table 1. The 
corresponding $\chi^{2}$ value is excellent~: $\chi^{2}/dof=0.86$ to be 
compared with the PDG value $\chi^{2}/dof=0.93$. 
The quality of the fit is illustrated in fig. 2.

The value $\varepsilon=0.132$ (see table 1) is certainly bigger than the 
DL value $\varepsilon=0.081$ or the PDG value $\varepsilon=0.093$ and it 
appears as being in between the effective Pomeron intercept value 1.1 
and the bare Pomeron intercept value 1.2 \cite{12}. However a direct 
comparison of different $\varepsilon$ values is not yet significative~: all 
the existing data other than $\sigma$ have first to be refitted by using 
a 2-component Pomeron amplitude.

Note that the residue of the non-leading Regge trajectory $Y_{1}^{\gamma
\gamma}$, numerically close to 0, is not well determined~: its weight in  
the minimization is negligible due to the low precision of the low energy
$\gamma\gamma$ data.

Let us also note that the value $0.303\cdot 10^{-2}$ of the 
$\delta$-parameter is perfectly compatible with  
the generalized vector-dominance model \cite{13}.

It is interesting to explore the relative importance of $X$ and $Z$ 
components in $\sigma$, by plotting the ratio $R$ (see fig. 3)
\begin{equation}
	R=\frac{Xs^{\varepsilon}}{Z}.
	\label{16}
\end{equation}
It can be seen from fig. 3 that the $X$-component acts like an 
asymptotic component. However the asymptoticity is clearly delayed~: 
$X$ dominates over the $Z$ component only in the TeV region. The only 
exception is the $Kp$ scattering, where the asymptoticity occurs 
already in the ISR region of energies.

By using the analytic forms (\ref{15}) and the values of the parameters 
given in table 1, one can make detailed predictions for $\sigma$ at 
high energies, in particular in the RHIC, LHC and cosmic-rays regions 
of energies - see table 2. However, one should not consider too 
seriously such predictions~: unitarization will certainly introduce 
important corrections at high energies. The $X$-component, as the DL 
Pomeron, violates unitarity.

In conclusion, we propose a QCD-inspired analytic form of the Pomeron, a
2-pole Pomeron form, which gives an excellent fit to the  $pp,\ \pi 
p,\ Kp,\ \gamma p$ and $\gamma\gamma$ total cross-sections. Compared to 
the PDG fit with a simple Pomeron-pole, our fit has a better $\chi^{2}/dof$
with a smaller number of parameters. This 2-pole Pomeron form has the
advantage to be fully compatible with the weak Regge exchange-degeneracy
, universality, Regge factorization and the generalized vector
dominance model.

The theoretical and phenomenological implications of the 2-compo\-nent 
Pomeron are important and therefore they should be explored in the 
future in a detailed way.

\vspace{1cm}

\noindent \textbf{Acknowledgements}. We thank Prof. Vladimir Ezhela 
for kindly putting to our disposal the PDG data and for very fruitful 
discussions.We also thank Prof. Alphonse Capella for useful remarks.

\newpage

$$
\begin{tabular}{|c|c|c|c|c|c|}
	\hline
	$\varepsilon$ & $\delta,10^{-2}$ & X & $Z_{pp}$ & $Z_{\pi p}$ 
	& $Z_{Kp}$  \\
	\hline
	0.132 & 0.303 & 7.572 & 20.251 & 5.283 & 2.208  \\
	\hline
\end{tabular}
$$

$$
\begin{tabular}{|c|c|c|c|c|c|c|c|}
	\hline
	$Y_{1}^{pp}$ & $Y_{2}^{pp}$ & $Y_{1}^{\pi p}$ &  $Y_{2}^{\pi p}$ &
	$Y_{1}^{Kp}$ & $Y_{2}^{Kp}$ & $Y_{1}^{\gamma p}$ & $Y_{1}^{\gamma\gamma}$
	   \\
	\hline
	74.811 & 29.918 & 48.972 & 6.028 & 34.483 & 11.935 & 0.121 & $\simeq 0 $  \\
	\hline
\end{tabular}
$$

\begin{center}
	Table 1~: The values of the parameters in the analytic forms 
	(\ref{15}).\\
	$\varepsilon$ and $\delta$ are pure numbers. The rest of 
	the parameters are in mb.
\end{center}

$$
\begin{tabular}{|r|l|l|l|l|l|l|l|l|}
	\hline
	$\sqrt{s},GeV$ & $\sigma_{\bar pp}$ & $\sigma_{pp}$ & $\sigma_{\pi^{+}p}$
	& $\sigma_{\pi^-p}$ & $\sigma_{K^{+}p}$& $\sigma_{K^-p}$
	& $\sigma_{\gamma p}$ & $\sigma_{\gamma\gamma}, 10^{-3}$  \\ 
	\hline
	100 & 46.8 & 46.3 & 31.4 & 31.3 & 28.2 & 28.0 & 0.140 & 0.421  \\
	\hline
	200 & 51.5 & 51.2 & 36.3 & 36.3 & 33.2 & 33.1 & 0.155 & 0.469  \\
	\hline
	300 & 54.8 & 54.7 & 39.7 & 39.7 & 36.6 & 36.6 & 0.166 & 0.501  \\
	\hline
	400 & 57.5 & 57.4 & 42.4 & 42.4 & 39.3 & 39.3 & 0.174 & 0.525  \\
	\hline
 	500 & 59.7 & 59.6 & 44.6 & 44.6 & 41.5 & 41.5 & 0.180 & 0.546  \\
	\hline
	600 & 61.6 & 61.5 & 46.6 & 46.6 & 43.5 & 43.5 & 0.186 & 0.564  \\
	\hline
	1800 & 75.4 & 75.4 & 60.4 & 60.4 & 57.3 & 57.3 & 0.228 & 0.692  \\
	\hline
        12000 & 111 & 111 & 96.4 & 96.4 & 93.3 & 93.3 & 0.337 & 1.02  \\
	\hline
	30000 & 136 & 136 & 121 & 121 & 118 & 118 & 0.413 & 1.25  \\
	\hline
\end{tabular}
$$
\begin{center}
	Table 2 : Extrapolation of the analytic forms (\ref{15}) at high 
	energies.\\
	$\sigma$ are given in mb.
\end{center}


\begin{thebibliography}{99}
	\bibitem{1}  I. Ya. Pomeranchuk, \textsl{Zh. Eksp. i Teor. Fiz.} 
	\textbf{34} 
	(1958) 725~; G.F. Chew and S.C. Frautschi, \textsl{Phys. Rev. Lett.} 
	\textbf{7} (1961) 394.

	\bibitem{2}  A. Donnachie and P.V. Landshoff, \textsl{Nucl. Phys.} 
	\textbf{B244} (1984) 322~;  A. Donnachie and P.V. Landshoff, talks
	at the Heidelberg Workshop \textsl{Pomeron and Odderon in Theory and
	Experiment}, March 1998.
	
	\bibitem{3}  Particle Data Group, Review of Particle 
	Physics, \textsl{Eur. Phys. J.} \textbf{C3} (1998) 205-212.

	\bibitem{4}  J.R. Cudell, V. Ezhela, K. Kang, S. Lugovsky and N. 
	Tkachenko, \textsl{Phys. Rev.} \textbf{D61} (2000) 034019.

	\bibitem{5}  Ref. 3, pp. 353-434.

	\bibitem{6}  S.P. Denisov \textsl{et al.}, \textsl{Phys. Lett.} 
	\textbf{B36} (1971) 415, 528~; \textsl{Nucl. Phys.} \textbf{B65} 
	(1973) 1;
	A.S Carroll \textsl{et al.}, \textsl{Phys. Lett.} 
	\textbf{B61} (1976) 303~; \textsl{Phys. Lett.} 
	\textbf{B80} (1979) 423.
	
	\bibitem{7}  P. Gauron, B. Nicolescu and O. Selyugin, \textsl{Phys. 
	Lett.} \textbf{B397} (1997) 305.
	
	\bibitem{8}  J.R. Cudell, K. Kang and S.K. Kim, \textsl{Phys. 
	Lett.} \textbf{B395} (1997) 311.

	\bibitem{9}  J. Bartels, L.N. Lipatov and G.P. Vacca, hep-ph/9912423.

	\bibitem{10}  L.G. Dakhno and V.A. Nikonov, \textsl{Eur. Phys. J.} 
	\textbf{A5} (1999) 209. 

	\bibitem{11}  A.R. White, hep-ph/0002303 and references quoted therein.

	\bibitem{12}  A. Capella, A. Kaidalov, C. Merino and J. Tran Thanh 
	Van, \textsl{Phys. Lett.} \textbf{B337} (1994) 358.
	
	\bibitem{13}  P. Ditsas, B.J. Read and G. Shaw,  \textsl{Nucl. Phys.} 
	\textbf{B99} (1975) 85~; P. Ditsas and G. Shaw,  \textsl{Nucl. Phys.} 
	\textbf{B113} (1976) 246~; J.J. Sakurai and D Schildknecht,  \textsl 
	{Phys. Lett.} \textbf{B40} (1972) 121. 
	
	\bibitem{14}  Computer readable data files are available at http://
	pdg.lbl.gov.
	
\end{thebibliography}
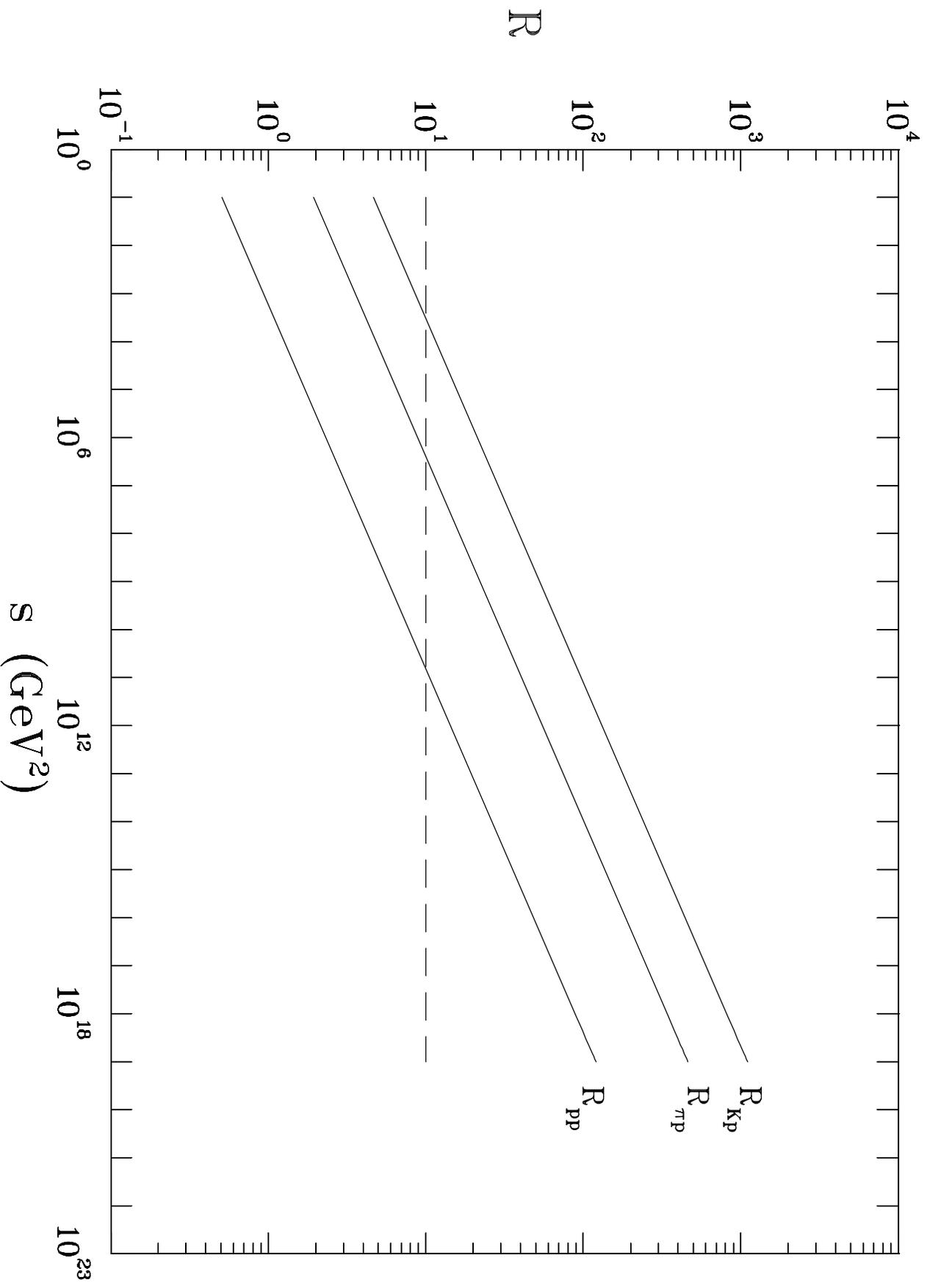
\end{document}